\documentclass[12pt]{elsarticle1}
\usepackage{amssymb,amsmath,amsfonts}
\usepackage{url}
\journal{arXiv}

\begin{document}
\begin{frontmatter}

\title{On the term of the 4-th order with respect to the field operators in the translation-invariant polaron theory}

\author[lak]{V.D.~Lakhno\corref{cor1}}
\ead{lak@impb.psn.ru}

\cortext[cor1]{Corresponding author}

\address[lak]{Institute of Mathematical Problems of Biology, Russian Academy of Sciences, Pushchino, Moscow Region, 142290, Russia}

\begin{abstract}
It is shown that 4-th order term in the translation-invariant polaron theory vanishes.
\end{abstract}


\end{frontmatter}

Having radically changed the concept of polarons, the theory of translation-invariant polarons (TI-polarons) ~\cite{1}-\cite{2} has recently came into focus of attention ~\cite{3}-\cite{8}. In this connection we discussed this theory in detail in review ~\cite{9}. Comments on papers ~\cite{3}-\cite{9}, that have come to the author suggest that most questions are concerned with vanishing of the contribution into the TI-polaron ground state energy made by the term of the 4-th order with respect to the field operators which arises in Froehlich Hamiltonian after Lee, Low, Pines (LLP) transformation ~\cite{LLP} (Appendix 1 in ~\cite{9}). Though the proof of this statement is given in ~\cite{1},~\cite{10} and in ~\cite{9}, it seems not to be explicit enough, some details are omitted. The aim of this paper is to discuss the point in detail.

According to ~\cite{9}, the term of the 4-th order with respect to the phonon field operators $H_1^{(4)}$ has the form:
\begin{equation}\label{1}
    H_1^{(4)} = \frac{1}{2m}\sum_{k,k'}\vec{k}\vec{k'}a_k^{+}a_{k'}^{+}a_ka_{k'}\,
\end{equation}
Accordingly, the contribution of the term $H_1^{(4)}$ into the ground state energy is:
\begin{equation}\label{2}
    E_1^{(4)} = \sum_{k,k'}\vec{k}\vec{k'} \rho_{\vec{k},\vec{k'}}\,
\end{equation}

\begin{equation*}
    \rho_{kk'} = \langle0|\Lambda_0^{+}a_k^{+}a_{k'}^{+}a_ka_{k'}\Lambda_0|0\rangle\,,
\end{equation*}

\begin{equation*}
    \Lambda_0 = C\exp\left(\frac{1}{2}\sum_{k,k'}a_k^{+}A_{kk'}a_{k'}^{+}\right),
\end{equation*}
where $A_{kk'}$ is a symmetrical matrix: $A_{kk'}=A_{k'k}$.
It is easy to see that:
\begin{equation}\label{3}
    a_{k'}\Lambda_0 = \sum_{k''}A_{k'k''}a_{k''}^{+}\Lambda_0
\end{equation}
Therefore:
\begin{equation}\label{4}
    \Lambda_{k,k'} = a_ka_{k'}\Lambda_0 = A_{k'k}\Lambda_0 + \sum_{k'',k'''}A_{k'k''}A_{kk'''}a_{k''}^{+}a_{k'''}^{+}\Lambda_0
\end{equation}

Hence, function $\rho_{\vec{k},\vec{k'}}$ in \eqref{2} is the norm of the vector $\Lambda_{k,k'}$:
\begin{equation}\label{5}
    \rho_{kk'} = \langle0|\Lambda_{kk'}^{+}\Lambda_{kk'}|0\rangle\,
\end{equation}
Let us show that the matrix $A_{kk'}$ has the structure:
\begin{equation}\label{6}
    A_{kk'} = (\vec{k}\vec{k'}) Q (|\vec{k}|,|\vec{k'}|) \,
\end{equation}
For this purpose let us use equation (7.7) from ~\cite{9} determining functional of the ground state $\Lambda_0$:
\begin{equation}\label{7}
\left(\sum_{k'}M_{1kk'}^{*}a_{k'}-\sum_{k'}M_{2kk'}^{*}a_{k'}^{+}\right)\Lambda_0|0\rangle = 0
\end{equation}
With the use of \eqref{3} and \eqref{7} we get the condition:
\begin{equation}\label{8}
\sum_{k'}M_{1kk'}^{*}A_{k''k'} - M_{2kk''}^{*} = 0
\end{equation}
According to ~\cite{1},~\cite{2}, matrix $M_{1,2kk'}$ has the structure: $M_{1,2kk'} = (\vec{k},\vec{k'}) R_{1,2} (|\vec{k}|,|\vec{k'}|)$. Hence, in accordance with condition \eqref{8} matrix $A$ \eqref{6} has the same structure. From \eqref{4}-\eqref{6} immediately follows that
\begin{equation}\label{9}
    \rho_{\vec{k},\vec{k'}} = \rho_{-\vec{k},\vec{k'}} = \rho_{\vec{k},-\vec{k'}}\,
\end{equation}
and $E_1^{(4)}$ \eqref{2} becomes zero which was to be proved.

Notice that if the total momentum of a TI-polaron $\vec{P}$ is nonzero, then matrix $A$ no longer has the structure of \eqref{6}: multiplier $Q$ in this case becomes angular dependent. Expression for the ground state energy $E(P)$ given in \cite{9} is valid in this case only in the limit $\vec{P}\to0$.

In conclusion the author would like to thank Prof.~Devreese for his recommendation to present proofs of \cite{1},~\cite{2},~\cite{9} in greater detail.

The work was supported by RFBR project N 13-07-00256.

\end{document}